\documentclass[pre,twocolumn,showpacs]{revtex4-1}
\usepackage{amsmath}
\usepackage{graphicx}
\usepackage{color}
\usepackage{amssymb}
\usepackage{wasysym}
\usepackage{epstopdf}
\input{epsf}
\usepackage{bm}
\usepackage{amsmath}
\usepackage{amssymb}

\usepackage{numcompress}

\begin{document}
\sloppy

\title{Dynamic Monte Carlo Simulations of Inhomogeneous Colloidal Suspensions}

\author{Fabi\'an A. Garc\'ia Daza$^1$}
\email{fabian.garciadaza@manchester.ac.uk}
\author{Alejandro Cuetos$^2$}
\author{Alessandro Patti$^1$}
\email{alessandro.patti@manchester.ac.uk}
\affiliation{$^1$Department of Chemical Engineering and Analytical Science, The University of Manchester, Manchester, M13 9PL, UK}
\affiliation{$^2$Department of Physical, Chemical and Natural Systems, Pablo de Olavide University, 41013 Sevilla, Spain}

\begin{abstract}
The Dynamic Monte Carlo (DMC) method is an established molecular simulation technique for the analysis of the dynamics in colloidal suspensions. An excellent alternative to Brownian Dynamics or Molecular Dynamics simulation, DMC is applicable to systems of spherical and/or anisotropic particles and to equilibrium or out-of-equilibrium processes. In this work, we present a theoretical and methodological framework to extend DMC to the study of heterogeneous systems, where the presence of an interface between coexisting phases introduces an additional element of complexity in determining the dynamic properties. In particular, we simulate a Lennard-Jones fluid at the liquid-vapor equilibrium and determine the diffusion coefficients in the bulk of each phase and across the interface. To test the validity of our DMC results, we also perform Brownian Dynamics simulations and unveil an excellent quantitative agreement between the two simulation techniques.

\end{abstract}

\pacs{82.70.Dd, 61.30.-v}

\maketitle


\section{Introduction}
\label{sec1}

Colloidal suspensions are two-phase systems where dispersed macromolecules, supramolecular aggregates or solid particles, whose characteristic length and time scales are within the \textit{nano} domain, are homogeneously immersed in a continuum medium. Colloidal suspensions, hereafter simply referred to as colloids, exhibit a very rich phase behaviour, which is dramatically determined by the equilibrium of the enthalpic and entropic contributions driven by the interactions established between their constituents. Similarly to atomic and molecular systems, colloids can form fluid and solid phases and can display phase coexistence. For instance, liquid-vapor coexistence has been reported in colloids of spherical particles \cite{Vrabec20061509,  Thol201525}, while liquid crystals coexisting with isotropic  phases have been reported in colloids of anisotropic particles under different conditions of density, temperature and particle architecture \cite{BOLHUIS1997, CUETOS20151137, PATTI_2018516}. The dynamics of colloidal particles, commonly referred to as Brownian motion, is governed by the stochastic collisions with the molecules of the surrounding fluid and can be efficiently described by Brownian Dynamics (BD) and Dynamic Monte Carlo (DMC) simulations. In BD simulations, the dynamics is driven by thermal fluctuations, and the temporal evolution of suspended particles is determined from the time integration of stochastic differential equations. By contrast, in DMC simulations, particles perform random walks in the phase space and their attempted moves are accepted according to a transition probability that satisfies the simple balance condition with respect to the Boltzmann distribution \cite{Patti2012}.

Understanding the dynamics of colloids is of crucial importance in a range of applications related, for instance, to the paint and printing industry and the formulation of personal-care products. In most practical cases, external forces, determining the system deformation and flow, and more than a single phase can be involved. For example, the presence of an interface affects the kinetics of formation of crystalline nuclei in metastable fluids \cite{espinosa2016}. Recent theoretical and molecular simulation works have proposed innovative methods to investigate the dynamics of colloids. These include the dynamic version of the Single Chain Mean Field theory, which can predict and explain the exchange kinetics of pluronics from micellar aggregates to water solutions \cite{GARCIA20176794, GARCIA2017PRL}, and the DMC simulation technique formulated to mimic equilibrium \cite{Patti2012, SANZ2010, SANZ2011, Cuetos2015} and out-of-equilibrium \cite{Corbett201815118} processes. Molecular Dynamics (MD) and BD simulations are generally considered the techniques of choice to study the dynamics of molecular and colloidal systems. However, while the deterministic dynamics of atoms and molecules is well grasped by MD, the Brownian motion of colloids can only be reproduced by MD simulations that explicitly incorporate both dispersed and continuous phases, namely colloidal particles and solvent molecules, obviously resulting in a significant increase of the degrees of freedom and an extremely high computational cost. By contrast, BD simulations are able to implicitly mimic the presence of the solvent by incorporating drag and stochastic forces that act on the colloidal particles. Both techniques face a similar challenge: the time-step to integrate the equations of motion must be relatively small to guarantee accuracy at the cost of achieving long time-scales. This is a minor problem in dilute systems whose structural relaxation decay is generally within few nanoseconds, but becomes a challenge in dense or arrested systems, whose dynamics fully develop over much longer time-scales.

The DMC simulation technique can circumvent these limitations. It basically follows the Metropolis acceptance criterion and, although the resulting stochastic evolution of states would be time-independent, in the limit of small displacements it becomes equivalent to BD simulations, acquiring a time-dependent identity \cite{Kikuchi1991335, Kotelyanskii19925383, Heyes1998447}. Over the last few years, a significant number of works has been published on the application of DMC as an alternative to BD to investigate the dynamics of dense colloids. Sanz and co-workers \cite{SANZ2010, SANZ2011} studied the diffusion and nucleation of a colloidal suspension of spherical and anisotropic particles. Jabbari-Farouji and Trizac \cite{SARA2012} evaluated the performance of DMC compared to BD by relating the short-time diffusion of MC simulation with the infinite-dilution diffusion coefficient for spherical and disk-like particles. The DMC method proposed by our group makes use of the acceptance rate, $\mathcal{A}$, of elementary moves to rescale the MC time step and demonstrate the existence of a unique MC time scale that allows for a direct comparison with BD simulations \cite{Patti2012, Cuetos2015}. Our results were in excellent quantitative agreement with BD simulations of isotropic and liquid crystal phases. More recently, we extended the DMC technique to investigate the dynamics during transitory unsteady states, where the thermodynamic equilibrium of a colloid is modified by an external field resulting in an out-of-equilibrium dynamics \cite{Corbett201815118}. In this case, the MC time step was rescaled with a time-dependent acceptance rate, $\mathcal{A}(t)$. 

Despite the increasing use of the DMC method to investigate the dynamics of complex fluids, including liquid crystals \cite{LEBOVKA2019, LEBOVKA20192, CHIAPPINI2020, CUETOS2020}, most of the attention has been devoted to develop and test the DMC technique in homogeneous, single-phase colloids. However, due to the fundamental and practical relevance of understanding the dynamics close and across an interface, a theoretical formulation of DMC for systems where phase coexistence is observed needs to be explored. The aim of the present work is providing the theoretical framework to extend our DMC method to the study of heterogeneous systems. To this end, we investigate the transport of Lennard-Jones particles at the vapor-liquid phase coexistence\cite{Vrabec20061509, Thol201525, Smit19928639, Trokhymchuk19998510} and calculate their diffusivities at different temperatures in the vapor and liquid phases and across their interface. To test the validity of our method, the diffusion coefficients have also been obtained by performing BD simulations, which exhibit an excellent quantitative agreement with the DMC results. 

This paper is organised as follows. In Section 2, the theoretical aspects of the DMC formalism for heterogeneous systems are presented. In Section 3, we provide details on the Lennard-Jones model employed and on the DMC and BD simulation protocols. In Section 4, we analyse our results highlighting the migration of spherical particles through the two-phase system by DMC and BD simulations. Finally, in the last section, we discuss the most important conclusions of our study.

\section{Theoretical Framework}
\label{sec2}
In this section, we extend the DMC formalism valid for single-phase colloids to multiphase systems. For simplicity, we will present our methodology for the specific case of a biphasic liquid-vapor system. Its generalisation to the case of coexistence between multiple phases is straightforward. We assume that the liquid and vapor phases are arranged in an elongated cuboidal box of dimensions $L_z>>L_x=L_y$. The orientation of the liquid/vapor interfaces is identified by a unit vector perpendicular to them and parallel to the $z$ direction. The cuboid is discretized into layers that are parallel to each other and to the interface, and whose thickness is $L_z/n_V$, where $n_V$ is an arbitrary number of layers. Each layer $i$ contains $\mathcal{N}_i$ particles and occupies a volume $\mathcal{V}_i$, so that the box volume is $V=\sum_{i=1}^{n_{V}}\mathcal{V}_i$ and the total number of particles is $N=\sum_{i=1}^{n_{V}}\mathcal{N}_i$. Clearly, a liquid-like layer hosts more particles than a vapor-like layer, but this difference is less significant for adjacent layers, especially if they are far from the interface.  

\begin{figure}[h!]
  \includegraphics[scale=0.75]{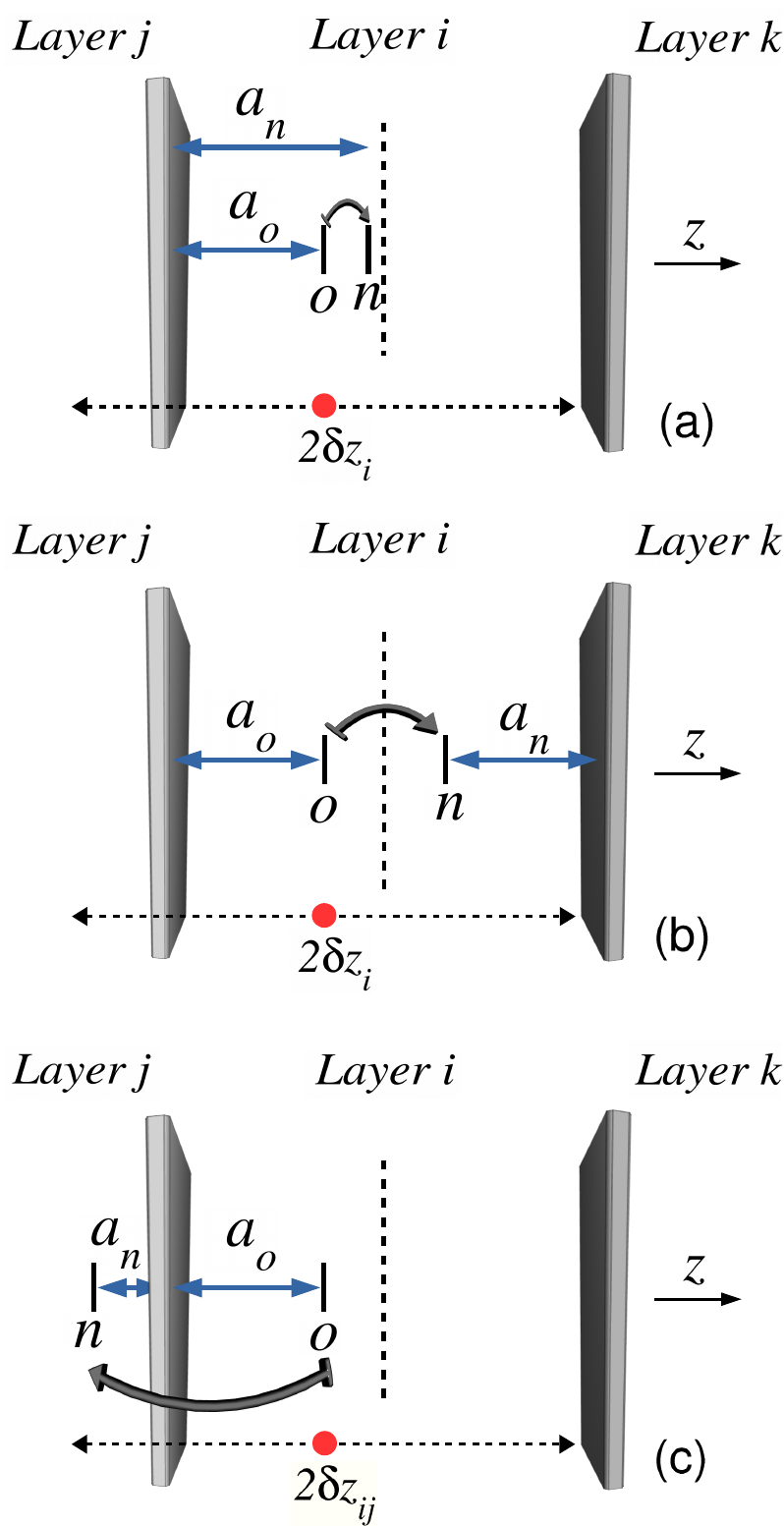}
  \centering
  \caption{Schematic representation of a particle performing an $o \rightarrow n$ move within the same layer (frames \textit{a} and \textit{b}) and to an adjacent layer (frame \textit{c}). The intervals $2\delta z_i$ and $2\delta z_{ij}$ indicate the allowed range of displacements for in-layer and inter-layer movements, respectively. The vertical dashed lines indicate the center of the layer. See text for details.
  \label{layer}}
\end{figure}

Following our previous works, we define an MC move as an attempt to simultaneously update the $f$ degrees of freedom of a randomly selected particle. In particular, a particle that belongs to layer $i$ can be moved to a generic new position $\xi=\left(\xi_{1,i},\xi_{2,i},\ldots,\xi_{f,i}\right)$ within an $f$-dimensional hyperprism of sides $\left[-\delta\xi_{\kappa,i},\delta\xi_{\kappa,i}\right]$ and volume $V_{\Xi ,i}=\prod_{\kappa=1}^{f}(2\delta\xi_{\kappa,i})$. Because the liquid phase is denser than the vapor phase, the vapor-like particles are more mobile than liquid-like particles. More generally, the maximum displacements in the unit of time, allowed to particles belonging to different layers, gradually change along the $z$ direction, from one phase to another. This implies that the length of the hyperprism sides, $2\delta\xi_{\kappa,i}$, is not the same across the $n_V$ layers. This is a key difference with respect to homogeneous systems, where the hyperprism volume can be assumed to be space-independent \cite{Patti2012, Cuetos2015}. Setting different displacements within the same simulation box can have an implication on the reversal moves of particles diffusing from a layer to another and might break the detailed balance. We have made sure that this is not the case by properly setting the volumes accessible to particles moving within the same layer or across different layers, as the probabilities calculated in Appendix A clarify. With reference to Fig.\ \ref{layer}, a particle originally located in layer $i$ can either move within the same layer or move to an adjacent layer. In the former case, the nearest adjacent layer to the particle might stay the same (Fig.\ \ref{layer}a) or change (Fig.\ \ref{layer}b). The
probability of accepting these trial in-layer and inter-layer moves is reported in Eq.~\ref{eq1} for an in-layer move with a change in the nearest adjacent layer, in Eq.~\ref{eq2} for an in-layer move with no change in the nearest adjacent layer, and in Eq.~\ref{eq3} for an inter-layer move.

\begin{widetext}
\begin{equation}
\label{eq1}
acc^{\text{in}(1)}_{i}=\min \left[ 1,e^{-\Delta U/k_BT}\left(\frac{\delta z_{ik}+m_1}{\delta z_{ij}+m_2}\right) \left( \frac{\delta z_{i}+m_3}{\delta z_{i}+m_4} \right) \frac{\delta z_{ij}}{\delta z_{ik}} \right]
\end{equation}

\begin{equation}
\label{eq2}
acc^{\text{in}(2)}_{i}=\min \left[1,e^{-\Delta U/k_BT} \left(\frac{\delta z_{ij}+m_0}{\delta z_{ij}+m_2}\right) \left( \frac{\delta z_{i}+m_3}{\delta z_{i}+m_4} \right) \right]
\end{equation}

\begin{equation}
 \label{eq3}
 acc^{\text{out}}_{i}=\min \left[1,e^{-\Delta U/k_BT} \right],
\end{equation}
\end{widetext}

\noindent where $\delta z_i$ is the maximum displacement along the $z$ direction in layer $i$, while $\delta z_{ij}=(\delta z_i+\delta z_j)/2$ and $\delta z_{ik}=(\delta z_i+\delta z_k)/2$. Additionally, $m_0=\min[a_n,\delta z_{ij}]$, $m_1=\min[a_n,\delta z_{ik}]$, $m_2=\min[a_o,\delta z_{ij}]$, $m_3=\min[a_o,\delta z_{i}]$ and $m_4=\min[a_n,\delta z_{i}]$. Finally, $a_o$ and $a_n$ are the distances of the particle from, respectively, the closest adjacent layer before and after the movement, $\Delta U$ is the energy change between the new and old configuration, $k_B$ the Boltzmann constant and $T$ the absolute temperature. A detailed derivation of Eqs.~\ref{eq1}-\ref{eq3} is available in Appendix A. It turns out that, as long as the density change across adjacent layers is smooth or, equivalently, the layers are relatively thin,  $acc_i^{\text{in}(1)}\approx acc_i^{\text{in}(2)}\approx acc_i^{\text{out}} \equiv acc_i = \min[1, \exp\left(-\Delta U/k_BT\right)]$. This result is exact if (\textit{i}) the particle move from a layer to another, or (\textit{ii}) original and adjacent layers are in the same bulk phase, but it is only approximate if the particle move across layers that are close or within the interface region (see Appendix A for details). The excellent agreement between DMC and BD simulation results shown in the next section suggests that this approximation is completely reasonable.

In the light of these considerations, the probability, $P^p_{\text{move},i}$, of moving a particle $p$ can be estimated regardless of whether $p$ actually remains in its original layer $i$ or not. However, $P^p_{\text{move},i}$ is still determined by the extension of the $f$-dimensional hyperprism and by the acceptance rate, which depend on the system density and thus on the region of the simulation box where $p$ is located. In particular, $P^p_{\text{move},i}$ is the product of the probability of randomly selecting one of the $N$ particles in the box, the probability of moving this particle to a point within the hyperprism of volume $V_{\Xi,i}$, and the probability of accepting the move, defined as $\mathcal{A}_i \equiv \langle acc_i \rangle$ (see Appendix B for details). Equivalently, $P^p_{\text{move},i}=\mathcal{A}_i/(NV_{\Xi,i})$. Therefore, the probability of moving each of the $N$ particles in an MC cycle, being a cycle equal to $N$ statistically independent attempts of moving a particle, reads:

\begin{equation}
   \label{eq4}
    P_{\text{move},i}=\frac{\mathcal{A}_i}{V_{\Xi,i}}.
\end{equation}

\noindent We can apply this probability to calculate the mean displacement and the mean square displacement of a given degree of freedom $\kappa$. The former reads $\langle\xi_{\kappa,i}\rangle=0$ as expected for the randomness of Brownian motion, while the latter takes the form 

\begin{equation}
  \label{eq5}
\langle\xi_{\kappa,i}^{2}\rangle=\int_{V_{\Xi,i}}\xi_{\kappa,i}^{2}P_{\text{move},i}\:d\xi_{\kappa,i}=\frac{\mathcal{A}_i\delta\xi_{\kappa,i}^{2}}{3}.
\end{equation}

\noindent In the case of $N$ particles performing $\mathcal{C}_\text{MC}$ cycles, Eq.~\ref{eq5} reads

\begin{equation}
  \label{eq6}
  \langle\xi_{\kappa,i}^{2}\rangle=\mathcal{C}_{\text{MC}}\frac{\mathcal{A}_i\delta\xi_{\kappa,i}^{2}}{3}.
\end{equation}

\noindent To relate the DMC simulation time step, $\delta t_{\text{MC}}$, to the time unit in a BD simulation, $t_{\text{BD}}$, we apply the Einstein relation, which is equivalent to the Langevin equation at long times \cite{einstein1956investigations}:
\begin{equation}
  \label{eq7}
\delta\xi_{\kappa,i}^2=2D_{\kappa}\delta t_{\text{MC},i},
\end{equation}
where $D_{\kappa}$ and $\delta t_{\text{MC},i}$ are, respectively, the diffusion coefficient at infinite dilution corresponding to the degree of freedom $\kappa$, and the time needed to perform an MC cycle in the volume $\mathcal{V}_i$ in the MC timescale. Combination of Eqs.~\ref{eq6} and \ref{eq7} results in:

\begin{equation}
  \label{eq8}
  \langle \xi_{\kappa,i}^2\rangle=\frac{2}{3}\mathcal{A}_i\mathcal{C}_{\text{MC}}D_{\kappa}\delta t_{\text{MC},i}.
\end{equation}

\noindent A combination of the previous equation with the Einstein relation for BD simulations $\langle\xi_{\kappa,i}^2\rangle=2D_{\kappa}t_{\text{BD}}$ leads to the following relation between the MC and BD timescales,

\begin{equation}
  \label{eq9}
  t_{\text{BD}}=\frac{\mathcal{A}_i}{3}\mathcal{C}_{\text{MC}}\delta t_{\text{MC},i}.
\end{equation}

\noindent This result indicates that particles belonging to different layers have an identical BD time scale, but not necessarily the same MC time step. This clearly differs from previous studies on homogeneous colloids, where a single space-independent time step was used for the entire volume delimiting the physical space. In multi-phase systems, time step and acceptance rate are expected to be uniform in the bulk regions of each phase, but this uniformity does not hold in regions close or at the interface. In particular, the acceptance rate changes such that the condition given in Eq.~\ref{eq9} is fulfilled for each of the layers. Generalizing the previous results to the finite set of volumes $\{\mathcal{V}_1,\mathcal{V}_2,\ldots,\mathcal{V}_{n_{V}}\}$, we obtain the following relation:    

\begin{equation}
  \label{eq10}
\mathcal{A}_1\delta t_{\text{MC},1}=\mathcal{A}_2\delta t_{\text{MC},2}=\ldots=\mathcal{A}_{n_{V}}\delta t_{\text{MC},n_{V}}.
\end{equation}

\noindent This equation provides a relation between the MC time steps in each layer and guarantee the existence of a unique BD time scale $\delta t_{\text{BD}}$, as established by Eq.~\ref{eq9}. These results can be extended to polydisperse or monodisperse multicomponent systems \cite{Cuetos2015} in equilibrium or out-of-equilibrium processes \cite{Corbett201815118}.


\section{Model and Simulation Methodology}
\label{sec3}
\begin{figure}[h!]
   \includegraphics[scale=0.8]{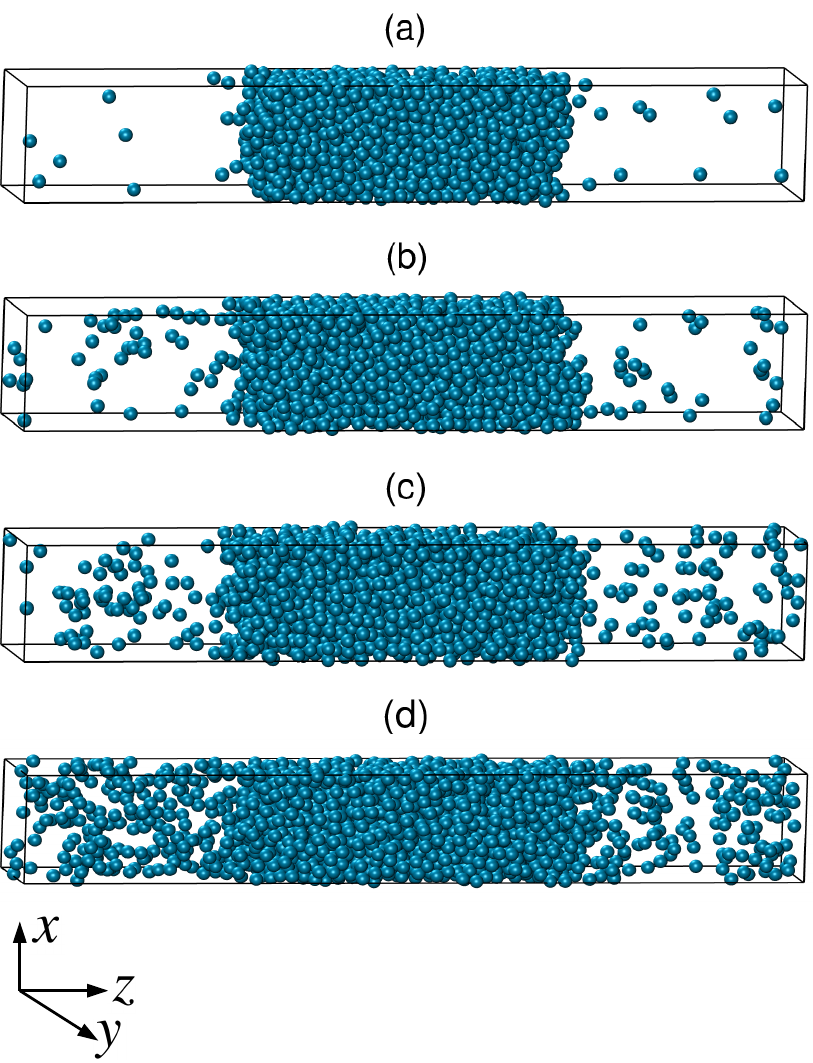}
  \centering
  \caption{Typical equilibrium configurations of LJ spheres forming liquid and vapor phases at $T^{*}=0.7$ (a), 0.8 (b), 0.9 (c) and 1.0 (d). 
  \label{fig2}}
\end{figure}

To test our theoretical formalism, we have run DMC and BD simulations of a system containing $N=2000$ spherical particles of diameter $\sigma$ in an elongated box with periodic boundaries and dimensions $L_x=L_y=10\sigma$ and $L_z=67.5\sigma$. The value of these parameters is within the range recommended for the study of liquid-vapor coexistence to avoid undesired finite size effects \cite{Chapela19771133, Holcomb1993437, Chen199510214}. The interaction between particles is determined by the Lennard-Jones (LJ) potential
\begin{equation}
  \label{eq11}
  U_{LJ}\left(r_{pq}\right)=4\epsilon\left[\left(\frac{\sigma}{r_{pq}}\right)^{12}-\left(\frac{\sigma}{r_{pq}}\right)^{6}\right],
\end{equation}
where $\epsilon$ is the depth of the potential and $r_{pq}$ is the separation distance between particles $p$ and $q$. The potential is truncated and shifted with a cut-off radius of $r_c=2.5\sigma$. Consequently, the potential actually employed in the simulations takes the form    
\begin{equation}
  \label{eq12}
  U\left(r_{pq}\right)=
  \begin{cases}
    U_{LJ}\left(r_{pq}\right)-U_{LJ}\left(r_c\right),& \text{if } r<r_c\\
    0,& \text{if } r\geq r_c.
  \end{cases}
\end{equation}
We use $\sigma$, $\epsilon$ and $\tau=\sigma^3\mu/\epsilon$ as the units of length, energy and time, respectively, where $\mu$ is the solvent viscosity. To equilibrate the systems, we performed standard MC simulations, each consisting of $10^7$ cycles, in the NVT ensemble at the reduced temperatures $T^*=k_BT/\epsilon=0.7,\:0.8,\:0.9$ and $1.0$. The initial configurations were prepared by arranging all the particles in a highly dense slab approximately located in the center of the simulation box along the $z$ direction. The systems were considered at equilibrium when the total energy and local densities reached steady values within reasonable statistical fluctuations. At this stage, a liquid and a vapor phase could be clearly identified, as shown in Fig.\ \ref{fig2}.

\subsection{DMC Simulations}
\label{sec3_1}
The simulation box was divided into $n_V=33$ planar  layers, each with a fixed volume $\mathcal{V}_i$ and parallel to the liquid-vapor interface, being perpendicular to the $z$ coordinate. In the light of this discretization, displacements, acceptance rates and MC time steps in Eqs.~\ref{eq7}-\ref{eq10} are subject to variations along the $z$ direction only. The DMC simulations have been performed in the NVT ensemble, with attempts to displace randomly selected particles according to the standard Metropolis algorithm. To properly reproduce the Brownian dynamics of colloidal spheres, no unphysical moves, such as swaps, jumps or cluster moves, have been performed. Particle displacements, $\delta\mathbf{r}$, are the sum of three orthogonal contributions oriented along the simulation box directions. More specifically,  $\delta\mathbf{r}=X_x\mathbf{\hat{x}} + X_y\mathbf{\hat{y}} + X_z\mathbf{\hat{z}}$, where $X_x$, $X_y$ and $X_z$ are random numbers that fulfill the condition $|X_l|\leq\delta r_{\text{max},i}$, where $\delta r_{\text{max},i}$ is the maximum displacement of the particle along a direction $l$ in $\mathcal{V}_i$. We stress that, in order to satisfy the detailed balance, the maximum displacements in the $z$ direction are subject to subtle modifications depending on whether a particle is displacing within the same layer or from a layer to another (see Appendix A). In particular, the maximum displacement in each layer $i$ depends on the particle diffusion coefficient at infinite dilution, $D_s$, and the arbitrarily set MC time step, $\delta t_{\text{MC},i}$, as obtained from the Einstein relation reported in Eq.~\ref{eq7}:  
\begin{equation}
  \label{eq13}
  \delta r_{\text{max},i}=\sqrt{2D_{s}\delta t_{\text{MC},i}},
\end{equation}
where, according to the Stokes-Einstein relation, $D_s\tau/\sigma^2=T^*/(3\pi)$. The dependence of $D_s$ on temperature is reported in Table \ref{tbl1}. The maximum displacement in every slab is therefore calculated by setting $\delta t_{\text{MC},i}$ and $D_s$.
\begin{table}[h]
  \caption{Dependence on temperature of diffusion coefficients at infinite dilution for spheres of diameter $\sigma$.}
   \label{tbl1}
  \centering
  \begin{tabular}{ccccc}
  \hline  
    $T^{*}$   & 0.7 & 0.8 & 0.9 & 1.0\\
    $D_s\tau/\sigma^2$ & 0.0743 & 0.0849 & 0.0955 & 0.1061\\
  \hline
  \end{tabular}
\end{table}
According to Eq.~\ref{eq10}, the MC time steps in all the layers are related to each other through their respective acceptance rates, which are not known \textit{a priori}. Since the full set of maximum displacements are not initially known, it is necessary to implement a preliminary trial-and-error procedure, which follows the algorithm described below:

\begin{enumerate}
\item Discretize the simulation box in a set of volumes $\{\mathcal{V}_1,\allowbreak\mathcal{V}_2,\allowbreak\ldots,\allowbreak\mathcal{V}_{n_V}\}$.

\item Set an arbitrary time step and a reference layer within the dense phase (i.e. $\delta t_{\text{MC},\text{ref}}$ in the central slab of the liquid phase).

\item Select a random particle, perform a random move and accept/reject it \textit{via} the  Metropolis algorithm.

\item Repeat this process $N$ times.

\item Calculate the the set $\{\mathcal{A}_i\}$ of the acceptance rates.

\item Keep $\delta t_{\text{MC},\text{ref}}$ fixed and update $\{\delta t_{\text{MC},i}\}$ \textit{via} $\delta t_{\text{MC},i}=\mathcal{A}_{\text{ref}}\delta t_{\text{MC},\text{ref}}/\mathcal{A}_{i}$.

\item Update the maximum displacement (more generally, the dimensions of the hyperprism) in each layer according to $\delta r_{\text{max},i}=\sqrt{2D_{s}\delta t_{\text{MC},i}}$.

\item Repeat from step 3 until  $\mathcal{A}_{1}\delta t_{\text{MC},1}=\mathcal{A}_{2}\delta t_{\text{MC},2}=\ldots=\mathcal{A}_{n_V}\delta t_{\text{MC},n_V}$.      
\end{enumerate}

In this work, the volumes $\{\mathcal{V}_i\}$ are defined by rectangular parallelepipeds with sides $L_x$, $L_y$ and $\delta L_z$. Additionally, we have considered three values for $\delta t_{\text{MC},\text{ref}}$, namely $10^{-3}\tau,\:10^{-2}\tau$ and $10^{-1}\tau$. The above algorithm exhibits a relatively fast convergence, which only takes a few thousand MC cycles, in agreement with previous DMC simulations  of mixtures of spherical and rod-like particles interacting via repulsive soft potentials \cite{Cuetos2015}. The so-obtained set of values of the maximum displacements, $\{\delta r_{\text{max},i}\}$, are finally used in the DMC production runs to generate the dynamic trajectories consistently with the particle mobility imposed by the local density. 


\subsection{BD Simulations}

In BD simulations, a stochastic differential equation, the so-called Langevin equation, is integrated forward in time and trajectories of particles are created \cite{LOWEN1994}. For spherical particles, the position of the center of mass of particle $p$ is defined as $\mathbf{r}_p$ and is updated in each BD step by the following equation:

\begin{equation}
\begin{split}
\label{eq14}
\mathbf{r}_p\left(t+\Delta t\right)=\mathbf{r}_p(t)+\frac{D_s}{k_BT}\mathbf{F}_p(t)\Delta t +\\ +
\left(2D_s\Delta t\right)^{1/2}\mathbf{R}_0(t).
\end{split}
\end{equation}

\noindent Here, $\mathbf{F}_p$ is the total force on particle $p$ exerted by the surrounding particles and $\mathbf{R}_0=(R_x,R_y,R_z)$ is a vector of Gaussian random numbers with unit variance and zero mean. In all BD simulations, the time step was set to $\Delta t=10^{-5} \tau$.   


\subsection{Comparison of Simulation Methodologies}

\par To compare DMC and BD simulation results, we analysed the particle dynamics in individual layers by measuring the long-time diffusion coefficients in each volume $\mathcal{V}_i$. More specifically, we calculated two-dimensional diffusion coefficients that refer to the motion of particles in $xy$ planes parallel to the liquid-vapor interface. These diffusion coefficients read

\begin{equation}
\label{eq15}
D_{xy}(\mathcal{V}_i)=\frac{\langle\Delta r_{xy}^2(t)\rangle_{\mathcal{V}_i}}{4t},
\end{equation}

\noindent where $\langle\Delta r_{xy}^2(t)\rangle_{\mathcal{V}_i}$ is the local mean square displacement (MSD) in the volume $\mathcal{V}_i$ defined as \cite{Liu20046595}
\begin{widetext}
\begin{equation}
\label{eq16}
\langle\Delta r_{xy}^2(t)\rangle_{\mathcal{V}_i}=\frac{1}{P(t)} \left< \frac{1}{N(0)}\sum_{i\:\in\:\lambda(0,t)} \left(r_{xy,i}(t)-r_{xy,i}(0)\right)^2 \right >,
\end{equation}
\end{widetext}
where $\lambda(0,t)$ is defined as the set of particles that reside in $\mathcal{V}_i$ within the time interval $[0,t]$, $N(0)$ the number of such particles at time $0$ and $N(0,t)$ the number of particles that remain in $\mathcal{V}_i$ up to time $t$. Particles leaving the volume $\mathcal{V}_i$ during this time interval are not counted. From these definitions, the survival probability can be calculated as:
\begin{equation}
\label{eq17}
P(t)=\left < \frac{N(0,t)}{N(0)} \right >.
\end{equation}

These properties have been measured to investigate the interface dynamics in a wide variety of systems, including water \cite{Liu20046595}, sodium chloride aqueous solutions \cite{Wick200515574}, supercritical $\text{CO}_2$ in ionic liquids \cite{Huang200517842}, liquid hydrocarbon molecules ($n$-octane) confined in inorganic $\alpha$-$\text{SiO}_2$ nanopores \cite{Wang201674}, and more recently glycerol confined in $\gamma$-$\text{Al}_2\text{O}_3$ nanopores \cite{camposvillalobos2019selfdiffusion}. In principle, one would need long time intervals to increase the accuracy of Eq.~\ref{eq15}, but the number of survival particles in a given layer dramatically decreases over time. If the layers were thicker, the survival probability would increase, but this would affect the ability of capturing the dynamics in the interface region. In general, a sensible selection of the sampling time and layer volume is key to obtain good statistics and achieve accurate results.


\section{Results and Discussion}
\label{sec4}

\begin{figure}[h!]
   \includegraphics[scale=0.9]{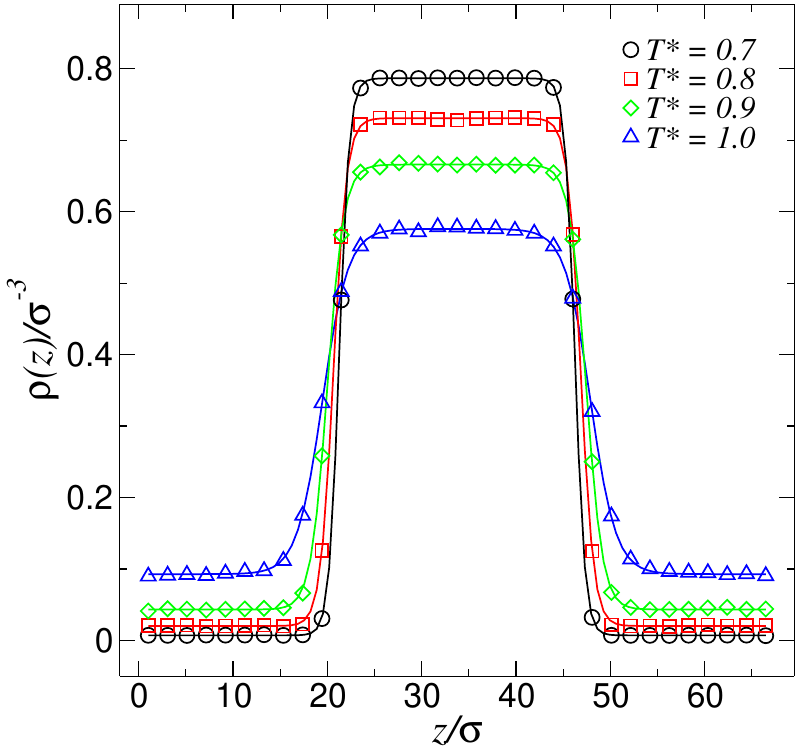}
  \centering
  \caption{Density profiles along the $z$ direction of the simulation box at $T^{*}=0.7$ (\textcolor{black}{$\ocircle$}), 0.8 (\textcolor{red}{$\square$}), 0.9 (\textcolor{green}{$\lozenge$}) and 1.0 (\textcolor{blue}{$\triangle$}). Symbols indicate DMC simulation results, whereas the solid lines are hyperbolic fits of the type given in Eq.~\ref{eq18}.}
  \label{fig3}
\end{figure}

To test the validity of our theoretical formalism, we have studied the dynamics of a system of 2000 spherical particles at $T^{*}=0.7,\:0.8,\:0.9$ and 1.0 by DMC and BD simulations. These temperatures are known to be lower than the critical temperature, which, for spheres interacting via a truncated and shifted LJ potential with the same cut-off radius as that employed in this work, ranges between 1.073 and 1.086 \cite{Vrabec20061509, Thol201525, Smit19928639, Trokhymchuk19998510,Haye1994556,  Dunikov20016623, Shi2001171}. The typical equilibrium configuration at these temperatures consists of a central liquid phase separated from a vapor phase by two roughly planar interfaces, as shown in Fig.~ \ref{fig2}. The thickness of the liquid region is approximately $25\sigma$ at $T^*=0.7$ and $28\sigma$ at $T^*=1.0$. Such a thickness, slightly larger than $10r_c$, guarantees that the central region of the liquid phase is not influenced by the interfaces. Upon increasing temperature, the phase diagram indicates that the density of the liquid and vapor phases must, respectively, decrease and increase. This is indeed what we observe at the end of our equilibration runs. In addition, we also notice a change in the thickness and density of the interfaces as the density profiles of Fig.~\ref{fig3} suggest. 

\begin{figure}[h!]
   \includegraphics[scale=0.9]{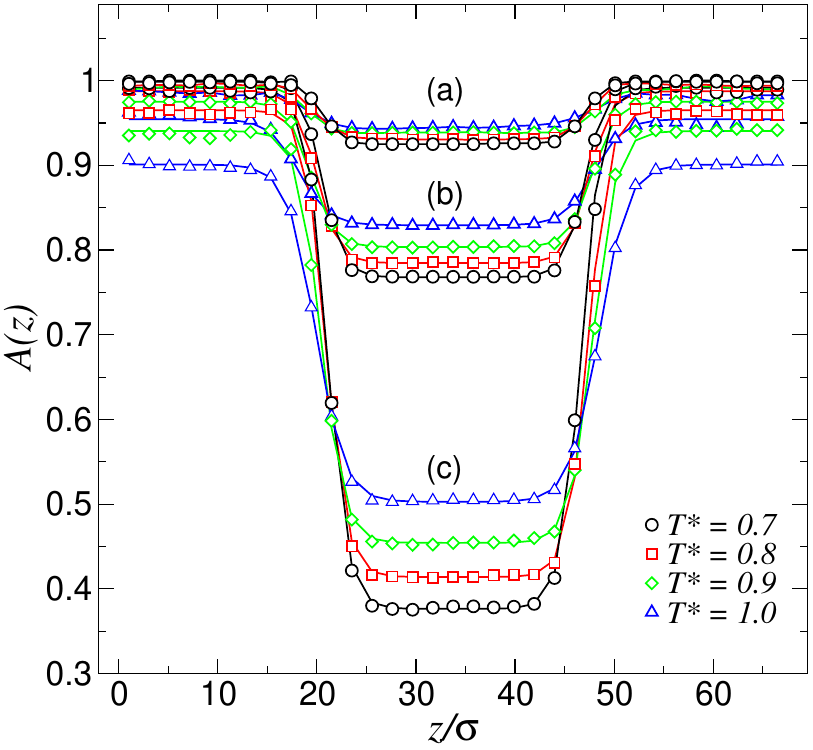}
  \centering
  \caption{Acceptance rates along the $z$ direction of the simulation box at three different values of the reference MC time step: (a) $\delta t_{\text{MC,ref}}/\tau=10^{-3}$, (b) $\delta t_{\text{MC,ref}}/\tau=10^{-2}$ and (c) $\delta t_{\text{MC,ref}}/\tau=10^{-1}$. The solid lines are hyperbolic fits of the type given in Eq.~\ref{eq18}.}
  \label{fig4}
\end{figure}

To obtain these profiles, we calculated the average number of particles, $\langle\mathcal{N}_z\rangle$, in each layer along the box $z$ direction and divided it by the layer volume $\mathcal{V}_z$, that is $\rho(z)=\langle\mathcal{N}_z\rangle/\mathcal{V}_z$. In Fig.~\ref{fig3}, we only report the density profiles as obtained with DMC simulations, being the corresponding BD simulation results basically identical and not shown here. In particular, the density of the bulk liquid phase, $\rho_{l,bulk}$, decreases from $0.788\sigma^{-3}$ at $T^*=0.7$ to $0.576\sigma^{-3}$ at  $T^*=1.0$, whereas the density of the bulk vapor phase, $\rho_{v,bulk}$, increases from $0.708\times 10^{-4}\sigma^{-3}$ to $0.092\sigma^{-3}$ at the same temperatures, in excellent agreement with former results \cite{Vrabec20061509, Thol201525, Trokhymchuk19998510, Haye1994556, Dunikov20016623, Shi2001171, Nijmeijer19883789, Adams19911383,  VanGiessen2009}. The density profiles shown in Fig.~\ref{fig3} have been fitted with a linear combination of hyperbolic tangent functions of the form \cite{Chapela19771133, Thompson1984530}:
\begin{equation}
\rho(z)=\frac{1}{2}\left(\rho_l+\rho_v\right)+\frac{1}{2}\left(\rho_l-\rho_v\right)\tanh\left[\frac{2(z-z_0)}{d}\right],
\label{eq18}
\end{equation}
where $z_0$ is the approximate location that is equidistant from the liquid and vapor phase, and $d$ is the interface thickness. In particular, $d/\sigma=1.94$, 2.43, 3.25 and 5.21 at $T^*=0.7$, 0.8, 0.9 and 1.0, respectively, in very good agreement with MD simulation results \cite{Vrabec20061509, VanGiessen2009}. The calculated values of the interface thickness are relevant to set a suitable thickness $\delta L_z$ of the layers of volume $\mathcal{V}_i$ that are needed for the discretization of the simulation box and the analysis of the dynamics. After some preliminary tests, we set $\delta L_z=2.04\sigma$, which is $0.1\sigma$ larger than the smallest interface thickness at the temperatures studied. Following this discretization, we applied the procedure described in Section 3.1 to calculate the acceptance rate in each layer, being reported in Fig.~\ref{fig4} for three different values of the reference time step: $\delta t_{\text{MC,ref}}/\tau=10^{-3}$, $10^{-2}$ and $10^{-1}$.   

\begin{figure}[h!]
   \includegraphics[scale=0.9]{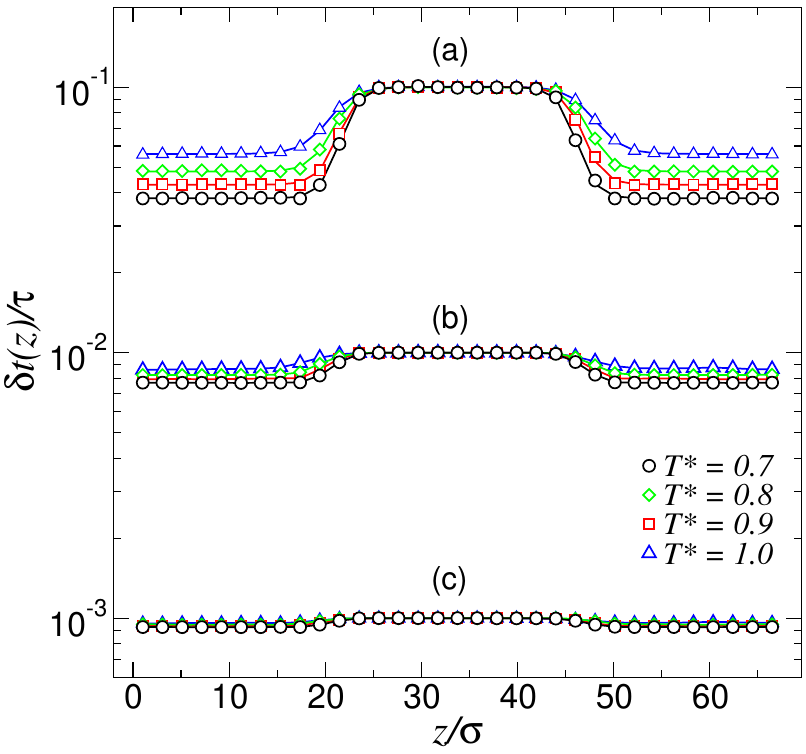}
  \centering
  \caption{MC time step along the $z$ direction of the simulation box for three different values of the reference MC time step: (a) $\delta t_{\text{MC,ref}}/\tau=10^{-1}$, (b) $\delta t_{\text{MC,ref}}/\tau=10^{-2}$ and (c) $\delta t_{\text{MC,ref}}/\tau=10^{-3}$. The solid lines are hyperbolic fits of the type given in Eq.~\ref{eq18}.}
  \label{fig5}
\end{figure}

In general, we observe an increase in the acceptance rate from the liquid, through the interface, to the vapor phase. The local acceptance rate is tightly related to the free volume available to the particles. Because this free volume is relatively small in the liquid phase, the probability of successfully moving liquid-like particles is lower than that of successfully moving vapor-like particles. This effect is especially evident at low temperatures, where the density difference between the liquid and vapor phase is larger (see Fig.\ \ref{fig3}). Particles at the interface display an intermediate behaviour, with an acceptance-rate profile that links the values of the bulk of each phase. More evident is the effect of the reference time step, $\delta t_{\text{MC,ref}}$. While the acceptance-rate difference between bulk liquid and vapor phase is up to 10\% at $\delta t_{\text{MC,ref}}/\tau=10^{-3}$ (0.92 \textit{vs} 1.0 at $T^{*}=0.7$), it increases to 60\% at $\delta t_{\text{MC,ref}}/\tau=10^{-1}$ (0.37 \textit{vs} 0.99 at $T^{*}=0.7$). This substantial disparity results from $\delta r_{\text{max}}\sim \sqrt{\delta t_{MC}}$ as increasing the MC time step leads to a remarkable increase of the maximum particle displacement and to a decrease in the acceptance rate. The rescaled MC time step in each layer, $\delta t_{\text{MC},i}$, is shown in Fig.~\ref{fig5}. 

\begin{figure}[h!]
   \includegraphics[scale=0.9]{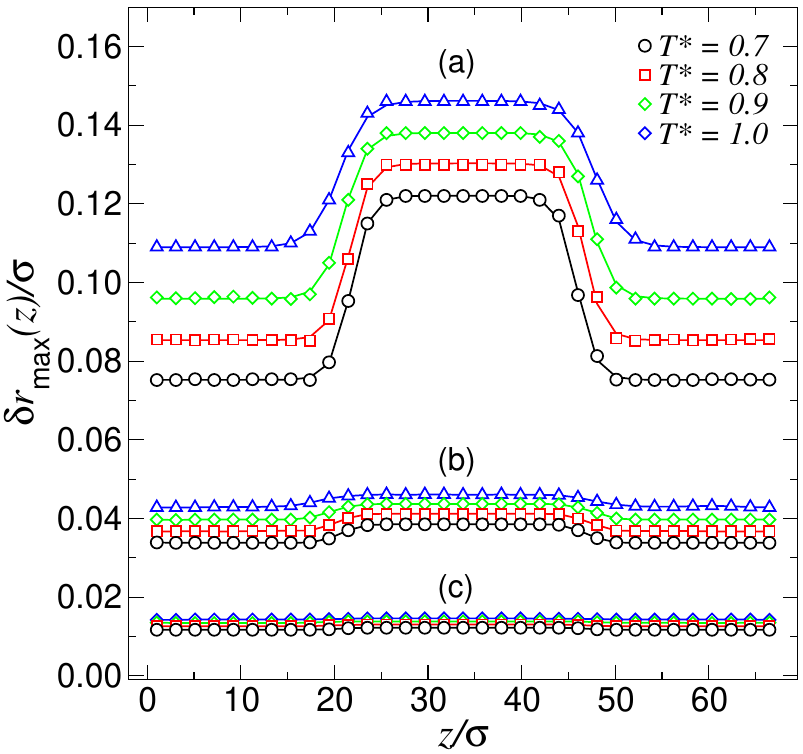}
  \centering
  \caption{Maximum displacements along the longitudinal direction of the simulation box for $\delta t_{\text{MC,ref}}/\tau$ equal to (a) $10^{-1}$, (b) $10^{-2}$ and (c) $10^{-3}$. The solid lines are hyperbolic fits of the type given in Eq.~\ref{eq18}.}
  \label{fig6}
\end{figure}

According to Eq.~\ref{eq10}, each of these time steps is proportional to the ratio $\mathcal{A}_{\text{ref}}/\mathcal{A}_{i}$. For layers that are especially close to the reference layer, which was arbitrarily set in the bulk liquid phase, $\mathcal{A}_{\text{ref}}/\mathcal{A}_i \approx 1.0$ and thus the associated MC time step is expected to be very similar to the reference time step. By contrast, for layers progressively farther from the reference layer, Fig.\ \ref{fig4} indicates that $\mathcal{A}_i>\mathcal{A}_{\text{ref}}$, resulting in a lower MC time step as one can observe in Fig.~\ref{fig5}. For this reason, $\delta t_{\text{MC},i}$ especially changes for $\delta t_{\text{MC,ref}}/\tau=10^{-1}$ (curves (a) in Fig.~\ref{fig5}), but less significantly as $\delta t_{\text{MC,ref}}/\tau$ decreases to $10^{-2}$ and $10^{-3}$ (curves (b) and (c) in Fig.~\ref{fig5}). Not surprisingly, the MC time step does not change with temperature in the liquid layers for each separate reference time step. This is simply due to the fact that $\delta t_{\text{MC,i}}=\delta t_{\text{MC,ref}}\mathcal{A}_{\text{ref}}/\mathcal{A}_i\approx\delta t_{\text{MC,ref}}$ in the bulk liquid and the value of $\delta t_{\text{MC,ref}}$ is the same for each set of curves (a), (b) and (c) in Fig.\ \ref{fig5}. By contrast, $\delta t_{\text{MC,i}} \ne \delta t_{\text{MC,ref}}$ in the interface or in the vapor phase, because $\mathcal{A}_i \ne \mathcal{A}_{ref}$. The corresponding maximum displacements, calculated from Eq.~\ref{eq13}, are reported in Fig.~\ref{fig6} for each layer along the longitudinal box direction. At a given $\delta t_{\text{MC,ref}}$, the maximum displacement increases with the particle diffusion coefficient at infinite dilution, which in turn increases with temperature (see Table \ref{tbl1}). We also notice that, upon decreasing $\delta t_{\text{MC,ref}}$, the difference between the maximum displacements across the layers reduces considerably, eventually disappearing at $\delta t_{\text{MC,ref}}/\tau < 10^{-3}$.

\par
\begin{figure}[h!]
   \includegraphics[scale=0.9]{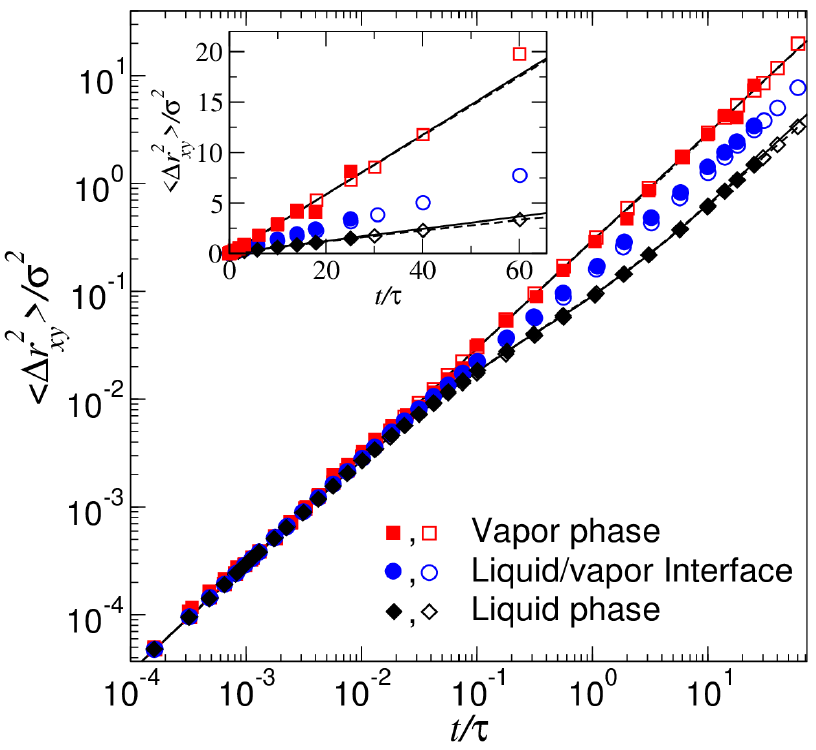}
  \centering
  \caption{MSDs of LJ particles calculated in liquid-like (\textcolor{black}{$\blacklozenge$,$\lozenge$}), vapor-like (\textcolor{red}{$\blacksquare$,$\square$}), and interface-like (\textcolor{blue}{$\CIRCLE$,$\ocircle$}) layers at $T^{*}=0.7$. Solid and empty symbols represent the MSDs obtained from BD and DMC simulations of heterogeneous systems, respectively. Solid and dashed lines refer to MSDs obtained from independent BD and DMC simulations of single phases, respectively. Note the double linear scale of the inset.}
  \label{fig7}
\end{figure}

 In the light of these preliminary results, which are essential to correctly mimic the Brownian dynamics and produce meaningful trajectories across the liquid-vapor interface, we now present the MSD and the diffusion coefficients in the bulk and at the interface. We stress that, within each layer, only the MSDs in the directions perpendicular to $z$ have been calculated here. Such two-dimensional MSDs are shown in Fig.~\ref{fig7} for $\delta t_{\text{MC,ref}}/\tau = 10^{-3}$ and $T^*=0.7$ (similar tendencies are observed at larger temperatures). The filled and empty symbols refer to the MSDs obtained from BD and DMC simulations, respectively, in liquid-like (diamonds), vapor-like (squares) and interface-like (circles) layers, whereas the solid and dashed lines are MSDs obtained from BD and DMC in single-phase systems. In particular, the DMC simulations of single-phase systems have followed the procedure reported in our previous work on monocomponent systems, with $\delta t_{\text{MC}}/\tau = 10^{-4}$ \cite{Patti2012}.

\begin{figure}[h!]
   \includegraphics[scale=0.9]{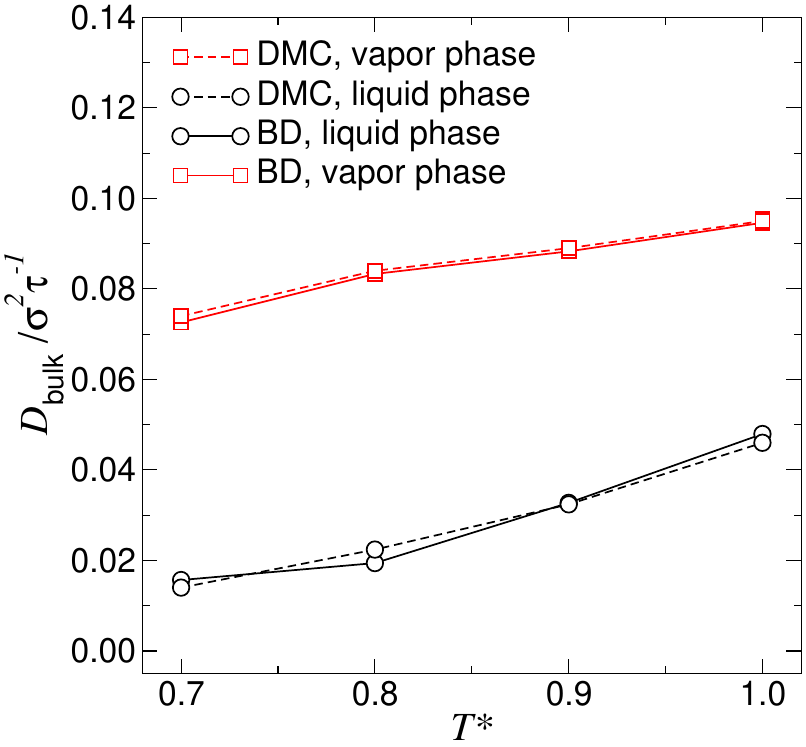}
  \centering
  \caption{Diffusion coefficients in the bulk of liquid (\textcolor{black}{$\ocircle$}) and vapor (\textcolor{red}{$\square$}) phases. Solid and dashed lines refer to BD and DMC simulations, respectively. Error bars are smaller than the symbols' size.}
  \label{fig8}
\end{figure}

\noindent The inset of Fig.\ \ref{fig7} presents the MSDs in double-linear scale to better appreciate the degree of agreement of BD and DMC simulations of homogeneous systems with their corresponding simulations of heterogeneous systems. The quantitative agreement is very good, with some deviations detected in the MSD of vapor-like particles, which are especially mobile and thus less prone to remain in the same layer for a long time, so reducing the statistical precision. Due to the necessarily small time-step and the limited number of particles observed especially in  vapour-like layers, the MSDs obtained with BD simulations in discretized systems have been calculated up to $t\approx 25 \tau$. Beyond this time, much longer simulations would be needed to compensate the statistical uncertainty. In any case, the diffusive regime is reached well before this time, with no effect on our analysis. More specifically, particles in the vapor phase enter the long-time diffuse regime very quickly, while those in the liquid phase experience an intermediate sub-diffusive regime, approximately three decades long, before entering the long-time diffusive regime at $t/\tau\approx10$. Particles at the interface exhibit an in-between MSD, seemingly closer to that of the liquid phase for the specific case of Fig.\ \ref{fig7}, although vapor-like MSDs have been observed in interface-like layers closer to the vapor phase. From the long-time slope of the MSD, we obtained the diffusion coefficients, $D_{\text{bulk}}$, in the bulk vapor and liquid phases, which are reported in Fig.~\ref{fig8} as a function of temperature. Again, the agreement between BD and DMC simulations is very good. 

\begin{figure}[h!]
   \includegraphics[scale=0.9]{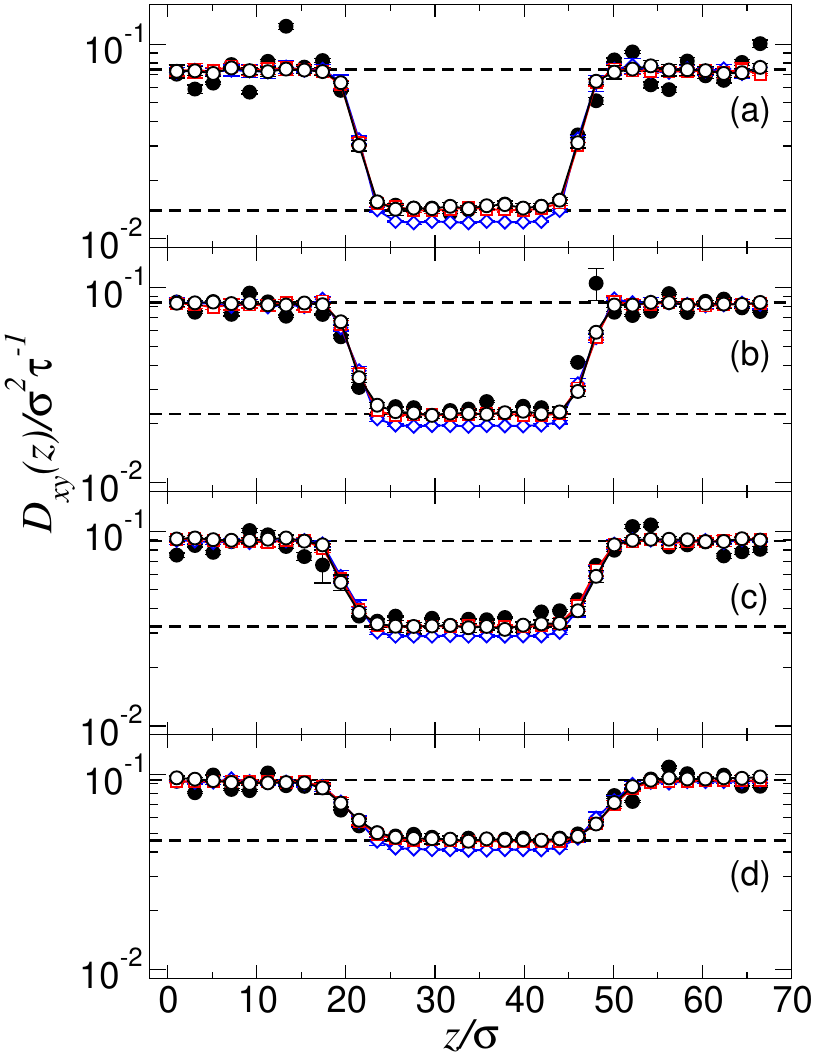}
  \centering
  \caption{ 
    Diffusion coefficients along the $z$ direction at (a) $T^{*}=0.7$, (b) $T^{*}=0.8$, (c) $T^{*}=0.9$ and (d) $T^{*}=1.0$, for $\delta t_{\text{MC,ref}}/\tau=10^{-3}$ (\textcolor{black}{$\ocircle$}), $10^{-2}$ (\textcolor{red}{$\square$}) and $10^{-1}$ (\textcolor{blue}{$\lozenge$}). Solid circles (\textcolor{black}{$\CIRCLE$}) refer to the diffusion coefficients obtained by BD simulations. Upper and lower dashed lines indicate the value of the diffusion coefficients calculated, respectively, in the vapor and liquid phase from DMC simulations in single-phase systems. The solid lines are a fit to a linear combination of hyperbolic functions of the type given in Eq.~\ref{eq18}.}
  \label{fig9}
\end{figure}

Having  established  that the DMC method is  able  to reproduce the Brownian dynamics in the bulk of the two coexisting phases, we now investigate the dynamics at the interface in more detail. To this end, we applied Eqs.\ \ref{eq16} and \ref{eq17} to calculate the diffusion coefficient, $D_{xy}(z)$, in the layers parallel to the liquid-vapor interface. Figure \ref{fig9} reports $D_{xy}(z)$ at different temperatures as calculated by BD and DMC simulations. Three different reference time steps have been considered in the DMC simulations, namely $\delta t_{\text{MC,ref}}/\tau = 10^{-3}$, $10^{-2}$ and $10^{-1}$. The diffusion coefficients in the liquid phase are consistently lower than those in the vapor phase, while those at the interface follow a trend that is well described by an hyperbolic function of the type given in Eq.\ \ref{eq18}. As temperature increases, the diffusion coefficients in both liquid and vapor phases increase too, closely following the DMC simulation results obtained in single-phase systems and reported as horizontal dashed lines. 

In general, the diffusive behaviour is well-captured by the DMC method at all the reference MC time steps. We notice that small fluctuations are observed for $D_{xy}(z)$ in the vapor phase, most likely due to the relatively small layer volume, $\{\mathcal{V}_i\}$, where the diffusion coefficients are calculated. A small $\{\mathcal{V}_i\}$ implies a short in-layer residence time, which determines the extent of the statistical noise in our results. By contrast, the statistical uncertainty associated to the $D_{xy}(z)$ measured in the liquid phase is negligible, since the in-layer residence time, in this case, is larger. We also notice that the DMC results are not influenced by the choice of $\delta t_{\text{MC,ref}}$ as all the rescaled results collapse on a single master curve. A modest discrepancy ($10\%$ to $14\%$) between the $D_{xy}(z)$ obtained by heterogeneous DMC and that obtained by single-phase DMC and BD simulations has been detected in the bulk liquid for $\delta t_{\text{MC,ref}}/\tau = 10^{-1}$. This difference is most likely due to the approximation  $acc^{\text{in}}\approx acc^{\text{out}}= \min[1, \exp\left(-\Delta U/k_BT\right)]$ discussed in Section \ref{sec2} and Appendix A. This assumption is fully satisfied at small $\delta t_{\text{MC,ref}}$, but becomes less reliable as $\delta t_{\text{MC,ref}}$ increases. For instance, in a liquid-like layer at $T^{*}=0.8$, $\mathcal{A}^{\text{in}}=0.936$ and $\mathcal{A}^{\text{out}}=0.930$ for $\delta t_{\text{MC,ref}}/\tau = 10^{-3}$ ($\approx 1\%$ difference), but $\mathcal{A}^{\text{in}}=0.468$ and $\mathcal{A}^{\text{out}}=0.417$ for $\delta t_{\text{MC,ref}}/\tau = 10^{-1}$ ($\approx 11\%$ difference). Assuming that $\mathcal{A}^{\text{in}}$ is smaller than its actual value leads to a shorter maximum displacement, which in turn limits the particle diffusion. Additionally, as discussed in Appendix A, the volume of the $f$-dimensional hyperprism near the layer border might change moderately in order to fulfill the reverse MC moves and preserve the detailed balance condition. If the hyperprism volume decreases, more particles will displace inside the layer and less to the adjacent layers. This can slightly increase the density of the liquid phase and thus slow down diffusion. The range of applicability of the DMC simulation technique is then set by the value of the reference time-step: a small $\delta t_{\text{MC,ref}}$ would imply demanding simulations to achieve the long-time dynamics, but exact results; while a large $\delta t_{\text{MC,ref}}$ would imply shorter simulations, but approximate results.


\section{Conclusions}
In summary, we extended the DMC simulation technique in order to investigate the dynamics of colloids that display phase coexistence. To this end, we first showed how to consistently displace particles that are in the bulk of each phase or in the interface region between them. More specifically, we discretized the space in thin layers $i$, assigned an arbitrary MC time-step to a reference layer, and then calculated the MC time-steps in the remaining layers by imposing the condition $\mathcal{A}_i\delta t_{\text{MC},i}=\mathcal{A}_{\text{ref}}\delta t_{\text{MC,ref}}$. This condition produces a specific maximum displacement for each layer that sets the limit of the actual particle displacement and ensures a consistent dynamics across the system. The probability of generating a new configuration, according to the detailed balance condition, follows an acceptance rule that depends on whether a particle attempts an in-layer or an inter-layer move. Nevertheless, in the limit of layers as thin as the interface, we have shown that the general Metropolis scheme with $acc_i = \min \left[ 1, e^{-\Delta U/k_BT} \right]$ can be applied regardless the specific movement being attempted. Finally, by rescaling the MC time-step with the acceptance rate, the results based on any arbitrary reference time-step collapse on a single master curve that reproduces the effective Brownian dynamics of the system. 

To test the validity of our theoretical framework, we studied a system of colloidal spheres interacting via a shifted and truncated LJ potential and displaying liquid-vapor coexistence in a range of temperatures. In particular, we calculated the MSD in the bulk liquid and vapor phases by running BD and DMC simulations in single-phase systems and compared it with the MSD obtained with the discretized BD and DMC methods in two-phase systems. The quantitative agreement between the four simulation methods was excellent, with the discretized DMC technique giving additional insight into the long-time dynamics at the interface between liquid and vapor. We also calculated the diffusion coefficients in the bulk and at the interface. Also in this case the agreement was very good, especially for shorter MC time-steps. Upon increasing the time-step, the acceptance rules for in-layer and inter-layer moves become more and more different. In this case, separate distribution probabilities should be applied or, to keep the efficiency of the method, the time-step should be reduced. 

The DMC methodology presented in this work is applicable to anisotropic particles with a higher number of degrees of freedom that display a rich phase coexistence behaviour. Nonetheless, there are some limitations that we would like to highlight here. First, the method can be applied to study steady-state processes, where the acceptance rate only changes in space, but it is constant over time. For out-of-equilibrium heterogeneous systems, space-time dependent acceptance rate distributions should be defined. Second, the DMC neglects the long-range solvent-mediated hydrodynamic interactions, which can have an impact in the dynamics of dense colloids. We also note that, in the current form, the present DMC methodology has been formulated for monodisperse systems, but it can be extended to multicomponent systems following the approach suggested in our previous work \cite{CUETOS2020}. 


\appendix 
\section{Detailed Balance in Dynamic Monte Carlo of Heterogeneous Systems}
\label{appenA}
\begin{figure}[h!]
  \includegraphics[scale=0.8]{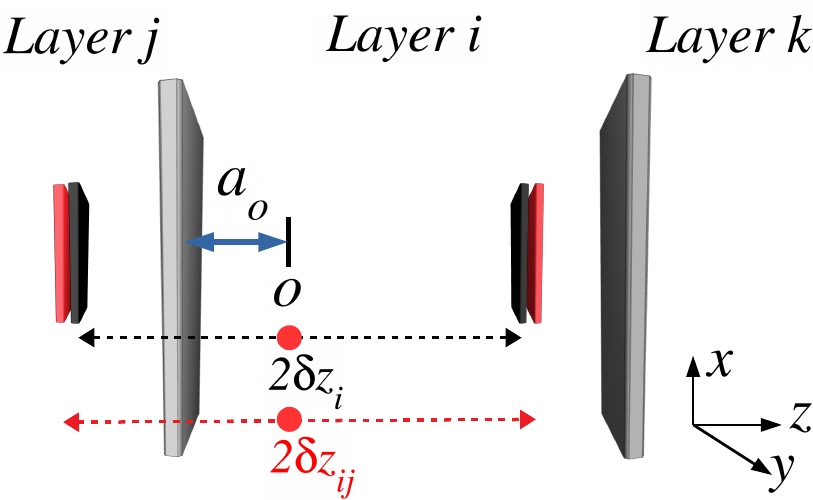}
  \centering
  \caption{A particle in layer $i$ can move within $i$ or to the adjacent layer $j$. In the former case, it will move within the interval $[-a_0, \delta z_{i}]$, in the latter case within the interval $[-\delta z_{ij},-a_0]$.}
  \label{appA1}
\end{figure}

Let us consider an elongated cuboidal box with square cross section in a $(x,y,z)$ Cartesian coordinate system, with its longest side oriented along $z$. This box can be assumed to be formed by parallel layers, piling on top of each other along $z$ and all with the same volume. Each layer contains a given number of particles that can perform in-layer or intra-layer displacements. Fig.~\ref{appA1} shows a particle initially located at position $o$ in layer $i$ and at a distance $a_o$ from its nearest adjacent layer $j$. 
This particle can move within an $f$-dimensional hyperprism delimited by $f$ degrees of freedom. For spherical particles, there are three degrees of freedom that define the hyperprism volume, which is $V_{\Xi,i}=8\delta x_i \delta y_i \delta z_i$, but only one is relevant in our discussion, namely the maximum displacement along $z$. If a particle originally in $a_0$ attempts an in-layer move, then its displacement is within the interval $[-a_0,\delta z_i]$. By contrast, if the particle attempts to move from layer $i$ to layer $j$, then the interval is $[-\delta z_{ij},-a_0]$, where $\delta z_{ij}=\left(\delta z_i+\delta z_j\right)/2$ is defined as the average of the maximum displacements that can be performed in contiguous layers. To meet the detailed-balance condition, the accessible volume must be defined in such a way that moves across layers are equally probable in both directions. To this end, we identified the nearest adjacent layer to which the particle can move (layer $j$, in Fig.~\ref{appA1}) and distinguished three possible scenarios: (1) a particle moves inside the same layer and its nearest adjacent layer changes; (2) a particle moves inside the same layer and its nearest adjacent layer does not change; and (3) a particle moves from a layer to another. In these three cases, the probability of performing a move is the product of four terms: the probability of finding a particle in a given configuration; the probability of moving this particle within the same layer or to an adjacent layer; the probability of selecting a new configuration; and the probability of accepting the move. In order to satisfy the detailed balance condition, we have defined the probability of moving a particle within the same layer as $(\delta z_{ij}+a_o)/2\delta z_{ij}$. Consequently, the probability of moving it to an adjacent layer reads $1- (\delta z_{ij}+a_o)/2\delta z_{ij}=(\delta z_{ij}-a_o)/2\delta z_{ij}$. The other terms are case-specific and are discussed below.


\subsection{In-layer displacements - the nearest adjacent layer changes}

\begin{figure}[h!]
   \includegraphics[scale=0.8]{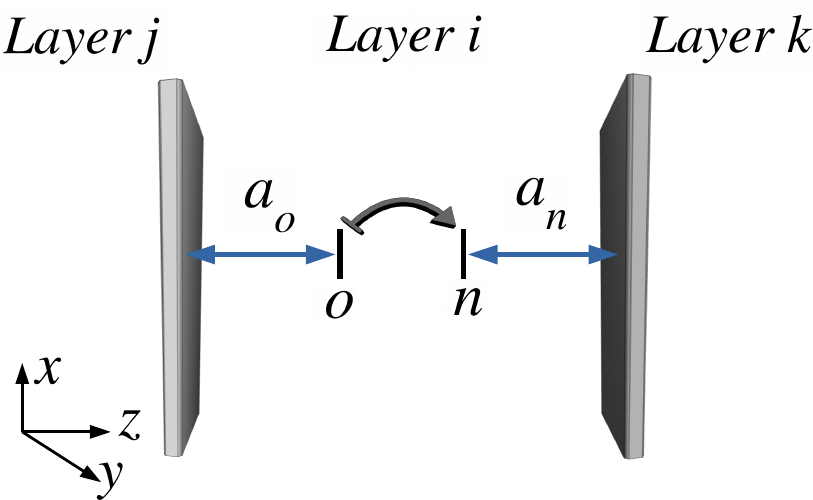}
  \centering
  \caption{A particle moving inside the same layer $i$ and changing its nearest adjacent layer from $j$ to $k$.}
  \label{appA2}
\end{figure}

This scenario is displayed in Fig.~\ref{appA2}. Initially, the particle is at position $o$ and at a distance $a_o$ from its nearest adjacent layer $j$. Following an MC move, its new nearest adjacent layer is $k$ and the distance from it is $a_n$.

\noindent To find the acceptance rule for this movement, we impose that the probability to move a particle from $o$ to  $n$ is the same as that of moving it from $n$ to $o$. More specifically,

\begin{equation}
  P_{on}=P_o \left(\frac{\delta z_{ij}+a_o}{2\delta z_{ij}}\right) \left(\frac{1}{\delta z_{i}+a_o}\right)acc_{i,on}^{\text{in}(1)},
\label{eqapp1}
\end{equation}

\noindent where $P_o$ is the probability of finding a particle in $o$; $\left(\delta z_{ij}+a_o\right)/2\delta z_{ij}$ is the probability of moving that particle within layer $i$; $1/\left(\delta z_{i}+a_o\right)$ is the probability of displacing the particle to a point in layer $i$; and $acc_{i,on}^{\text{in}(1)}$ is the probability of accepting the movement. Similarly,

\begin{equation}
  P_{no}=P_n\left(\frac{\delta z_{ik}+a_n}{2\delta z_{ik}}\right)\left(\frac{1}{\delta z_{i}+a_n}\right)acc_{i,no}^{\text{in}(1)}.
\label{eqapp2}
\end{equation}

\noindent To satisfy the detailed-balance condition, it is crucial to calculate the probability of moving a particle within the same layer using the average maximum displacements across contiguous layers, that is $\delta z_{ij}=\left(\delta z_i+\delta z_j\right)/2$ for $P_{on}$ and $\delta z_{ik}=\left(\delta z_i+\delta z_k\right)/2$ for $P_{no}$. As mentioned above, the two probabilities must be equal:

\begin{equation}
\begin{split}
  P_o\left(\frac{\delta z_{ij}+a_o}{2\delta z_{ij}}\right)\left(\frac{1}{\delta z_{i}+a_o}\right)acc_{i,on}^{\text{in}(1)}= \\ =P_n\left(\frac{\delta z_{ik}+a_n}{2\delta z_{ik}}\right)\left(\frac{1}{\delta z_{i}+a_n}\right)acc_{i,no}^{\text{in}(1)}.
  \label{eqapp3}
  \end{split}
\end{equation}

\noindent Therefore, the acceptance rule reads

\begin{equation}
\begin{split}
  acc^{\text{in}(1)}_{i}=\min \Big[1,e^{-\Delta U/k_BT} \cdot \\
 \cdot \left(\frac{\delta z_{ik}+a_n}{\delta z_{ij}+a_o}\right)\left(\frac{\delta z_{i}+a_o}{\delta z_{i}+a_n}\right)\frac{\delta z_{ij}}{\delta z_{ik}} \Big].
  \label{eqapp4}
  \end{split}
\end{equation}

\noindent If $a_o$ and $a_n$ are larger than the maximum displacement allowed, in order to ensure the correct normalization of the above probabilities, the acceptance rule should take the following general form

\begin{equation}
\begin{split}
  acc^{\text{in}(1)}_{i}=\min\Big[1,e^{-\Delta U/k_BT}\cdot \\
  \cdot \left(\frac{\delta z_{ik}+m_1}{\delta z_{ij}+m_2}\right)\left(\frac{\delta z_{i}+m_3}{\delta z_{i}+m_4}\right)\frac{\delta z_{ij}}{\delta z_{ik}}\Big],
  \label{eqapp5}
  \end{split}
\end{equation}

\noindent where $m_1=\min[a_n,\delta z_{ik}]$, $m_2=\min[a_o,\delta z_{ij}]$, $m_3=\min[a_o,\delta z_{i}]$ and $m_4=\min[a_n,\delta z_{i}]$ and the difference $\Delta U$ is the change in energy between the new and old states as a result of the MC move. 
\subsection{In-layer displacements - the nearest adjacent layer does not change}

In this case, $\delta z_{ik}=\delta z_{ij}$ and $a_n$ is the distance of the particle from the same nearest adjacent layer $j$ after the movement. Hence, the acceptance rule reads

\begin{equation}
\begin{split}
  acc^{\text{in}(2)}_{i}=\min \Big[1,e^{-\Delta U/k_BT}\cdot \\
  \cdot \left(\frac{\delta z_{ij}+m_0}{\delta z_{ij}+m_2}\right)\left(\frac{\delta z_{i}+m_3}{\delta z_{i}+m_4}\right) \Big], 
  \label{eqapp6}
  \end{split}
\end{equation}

\noindent where $m_0=\min[a_n,\delta z_{ij}]$.
\subsection{Inter-layer displacements}
\begin{figure}[h!]
   \includegraphics[scale=0.8]{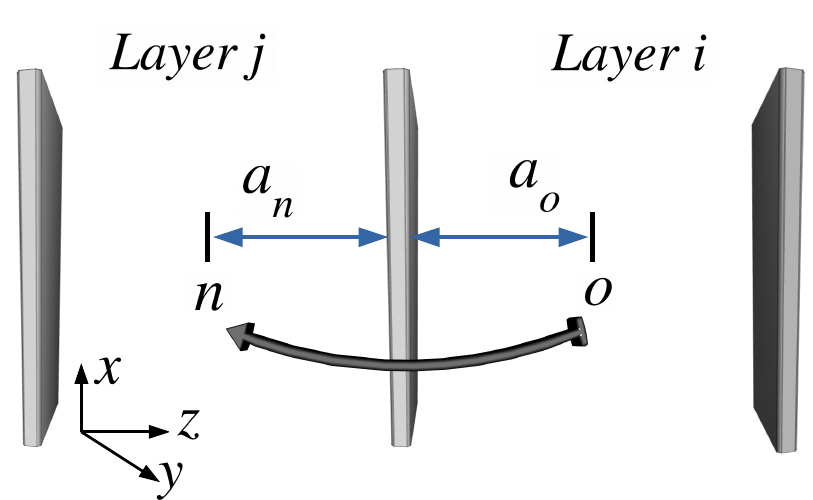}
   \centering
   \caption{Representation of a particle moving from an initial layer $i$ to and adjacent layer $j$.\label{appA3}}
\end{figure}

This case is illustrated in Fig.~\ref{appA3}. Using the previous definitions and similar arguments, the detailed balance condition now reads

\begin{equation}
\begin{split}
  P_o\left(\frac{\delta z_{ij}-a_o}{2\delta z_{ij}}\right)\left(\frac{1}{\delta z_{ij}-a_o}\right)acc^{\text{out}}_{i,on}=\\
  =P_n\left(\frac{\delta z_{ij}-a_n}{2\delta z_{ij}}\right)\left(\frac{1}{\delta z_{ij}-a_n}\right)acc^{\text{out}}_{i,no},
  \label{eqapp7}
\end{split}
\end{equation}

\noindent which, by using the Boltzmann distribution for the state probabilities, reduces to the general Metropolis criterion:

\begin{equation}
  acc^{\text{out}}_{i}=\min\left[1,e^{-\Delta U/k_BT}\right]. 
  \label{eqapp8}
\end{equation}

\noindent Finally, to test the validity of the three acceptance criteria, we consider the case of homogeneous systems. In this case, the maximum displacements remain constant regardless the layer, i.e. $\delta z=\delta z_i=\delta z_j=\delta z_k$, therefore, $\delta z=\delta z_{ij}=\delta z_{ik}$. In view of these considerations, the acceptance criteria $acc^{\text{in}(1)}_i=acc^{\text{in}(2)}_i=acc^{\text{out}}_i$, and thus the general Metropolis rule of Eq.~\ref{eqapp8} is recovered. For heterogeneous systems, if the layers are sufficiently small such that the changes in density are smooth in space, namely, the hyperprism volume does not change significantly across the layers, it is plausible to assume that the average maximum displacement in adjacent layers is basically constant, i.e. $\delta z_i\approx\delta z_{ij}\approx\delta z_{ik}$ and the acceptance probabilities  $acc^{\text{in}(1)}_i \approx acc^{\text{in}(2)}_i \approx acc^{\text{out}}_i = \min\left[1,e^{-\Delta U/k_BT}\right]$. We have employed this approximation in the DMC simulations discussed in this work.


\section{Mean Squared Displacement}
\label{appenB}
If the layer thickness is only slightly larger than the maximum particle displacement $\delta z_{\text{max}}$, the latter should change very smoothly from one phase to the other. Therefore, one can assume $\delta z_{ij}\approx\delta z_{ik}$, and $acc^{\text{in}(1)}_i=acc^{\text{in}(2)}_i=acc^{\text{in}}_i$ from Eqs.~\ref{eqapp5} and \ref{eqapp6}. Under such an assumption, the in-layer displacements can be computed with probability $\mathcal{A}_i^{\text{in}} \equiv \langle acc^{\text{in}}_i \rangle$, whereas the inter-layer displacements with probability $\mathcal{A}^{\text{out}}_i \equiv \langle acc^{\text{out}}_i \rangle$. In the light of these considerations and following the sketch of Fig.~\ref{appA1}, the mean displacement can be calculated as:

\begin{equation}
  \langle z_i\rangle=\int_{-a_o}^{\delta z_i} z\:P_{\text{move},i}^{\text{in}}\:dz+\int_{-\delta z_{ij}}^{-a_o}z\:P_{\text{move},i}^{\text{out}}\:dz,
\end{equation}

\noindent where $P_{\text{move},i}^{\:in}$ and $P_{\text{move},i}^{\:out}$ are the probabilities of moving inside and outside the layer $i$ within the hyperprism of volume $V^{\text{in}}_{\Xi,i}\propto[-a_o,\delta z_i]$ and $V^{\text{out}}_{\Xi,i}\propto[-\delta z_{ij},-a_o]$, respectively. As defined in Section~\ref{sec2}, these probabilities include the probability for the particle to stay or leave the layer, the probability of moving the particle in the region defined by $V^{\text{in}}_{\Xi,i}$ or $V^{\text{out}}_{\Xi,i}$, and the probability of accepting the move. Thereby, the mean displacement now reads

\begin{equation}
\begin{split}
    \langle z_i\rangle=\int_{-a_o}^{\delta z_i}\frac{\delta z_{ij}+a_o}{2\delta z_{ij}} \frac{\mathcal{A}^{\text{in}}_i}{\delta z_i+a_o}z\:dz+\\
    +\int_{-\delta z_{ij}}^{-a_o}\frac{\delta z_{ij}-a_o}{2\delta z_{ij}}\frac{\mathcal{A}^{\text{out}}_i}{\delta z_{ij}-a_o}z\:dz.
    \end{split}
\end{equation}

\noindent Contrary to what observed in homogeneous systems, this results indicates that $\langle z_i\rangle \ne 0$. However, if $\delta z_{ij}\approx\delta z_{i}$ and $ acc^{\text{in}}_i \approx acc^{\text{out}}_i$, then  $\mathcal{A}^{\text{in}}_i\approx \mathcal{A}^{\text{out}}_i$ and thus $\langle z_i\rangle \approx 0$. These conditions are fully accomplished in the bulk layers and also in the interface layers if these are sufficiently small. Analogously, the MSD can be calculated from

\begin{equation}
  \langle z_i^2\rangle=\int_{-a_o}^{\delta z_i}z^2P_{\text{move},i}^{\text{in}}dz+\int_{-\delta z_{ij}}^{-a_o}z^2P_{\text{move},i}^{\text{out}}dz.
\end{equation}

\noindent If the particle is restricted to move exclusively in the layer ($a_o>\delta z_{ij}$), only the first term in the right hand side becomes relevant and the MSD is equivalent to the one calculated in the case of single-phase systems. On the other hand, if the probability for the particle to move to its nearest layer is non zero and if we consider sufficiently small layers, such that $\delta z_{ij}\approx \delta z_{i}$, the MSD can be approximated by 

\begin{equation}
  \langle z_i^2\rangle = \frac{\mathcal{A}_i^{\text{in}}}{2\delta z_{i}}\frac{\delta z_i^3+a^3_0}{3}+\frac{\mathcal{A}_i^{\text{out}}}{2\delta z_{i}}\frac{\delta z_i^3-a^3_0}{3}.
\end{equation}

\noindent If we now incorporate the result of Appendix A, that is $\mathcal{A}^{\text{in}}_i \approx \mathcal{A}^{\text{out}}_i$, then

\begin{equation}
  \langle z_i^2\rangle=\frac{\mathcal{A}_i\delta z_{i}^2}{3}.        
\end{equation}
%
\section*{Declaration of Competing Interest}
 The authors declare that they have no known competing financial interests or personal relationships that could have appeared to influence the work reported in this paper.
%
\section*{Acknowledgements}
FAGD and AP acknowledge the Leverhulme Trust Research Project Grant RPG-2018-415 and the assistance given by IT services and the use of Computational Shared Facility at the University of Manchester. AC acknowledges the Spanish Ministerio de Ciencia, Innovaci\'on y Universidades and FEDER for funding (project PGC2018-097151-B-I00) and C3UPO for the HPC facilities provided.\\


\bibliography{main}

\end{document}